\documentclass{appolb}
\usepackage{graphicx}
\usepackage{amssymb}
\usepackage[intlimits]{amsmath}
\usepackage{amsfonts}
\usepackage{dsfont}
\usepackage{multirow}
\usepackage{subfigure}

\newcommand{\sh}[1]{#1\hskip-7pt \diagup}

\def\i{\ensuremath{\mathrm{i}}}

\def\T{\ensuremath{\mathrm{T}}}

\begin{document}

\title{Non-contact interactions and the hadronic light-by-light contribution to the muon $g-2$
\thanks{Excited QCD -- Sarajevo 3-9 February, 2013. Presented by R. Williams}
}
\author{Richard Williams$^1$, Christian S. Fischer$^2$, Tobias Goecke$^2$
\address{$^1$\small Institut f\"ur Physik, Karl-Franzens Universit\"at Graz, Graz, Austria\\
$^2$\small Institut f\"ur Theoretische Physik, Universit\"at Giessen, Giessen, Germany}
}
\maketitle
\begin{abstract}
We summarise recent results for the quark loop part of the light-by-light scattering
contribution to the muons anomalous magnetic moment. In particular
we focus on the impact of a momentum dependent quark and quark-photon vertex. We compare the  Dyson-Schwinger
description with that of the extended Nambu--Jona-Lasinio model (ENJL) and find
important quantitative differences. In particular the transverse parts of 
the quark-photon-vertex, which serve as a dynamical extension of simple vector meson dominance models,
do not yield the large suppression as found in the ENJL model.
\end{abstract}

\section{Introduction}

Here we give a brief summary of our results on the anomalous magnetic moment of
muon~\cite{Goecke:2012qm,Goecke:2011pe,Goecke:2013fpa}. This quantity serves to
provide a precision test of the standard model, in particular the electromagnetic
~\cite{Aoyama:2012wk}, weak~\cite{Czarnecki:2002nt} and strong force. The QED
contributions are dominant followed by QCD, with the latter dominating the 
theoretical error.

These QCD corrections are the leading order hadronic vacuum polarisation (LOHVP)~\cite{Hagiwara:2011af}
and the hadronic light-by-light scattering contribution (HLBL). The former may be inferred from
experiment~\cite{Jegerlehner:2009ry}, with Lattice results also becoming 
competitive~\cite{Feng:2011zk,Boyle:2011hu,DellaMorte:2011aa}. The HLBL contribution can only be
determined from theory, with many models attempting its evaluation
~\cite{Bijnens:1995xf,Hayakawa:1995ps,Knecht:2001qf,Melnikov:2003xd,Dorokhov:2008pw,Dorokhov:2012qa,Greynat:2012ww,Cappiello:2010uy,Fischer:2010iz,Goecke:2010if,Goecke:2012qm}. We focus here on two
approaches: the Extended Nambu--Jona-Lasinio (ENJL)  model~\cite{Bijnens:1995xf}; and the 
Dyson-Schwinger equations (DSEs)~\cite{Fischer:2010iz,Goecke:2010if,Goecke:2012qm}.

The combined theory result stands at $116\,591\,827.0\,(64)\times 10^{-11}$\cite{Prades:2009tw}, which compares to the experimental result of  $116\,592\,089\,(63)\times 10^{-11}$~\cite{Bennett:2006fi,Roberts:2010cj}. The discrepancy stands at $3\sigma$, but may be as high as $\sim 4.8\sigma$~\cite{Benayoun:2012wc}.

One interpretation of this discrepancy may be a signal of beyond the standard model physics. 
In these proceedings we argue a note of caution should be taken with the present HLBL estimation and its error. We believe that the limitations of models thus far used lead to an overly optimistic value. To demonstrate, we compare our approach with the ENJL
and highlight the differences and consequences in the
quark-loop contribution,. In particular, we show that the large suppression due
to VMD in the ENJL~\cite{Bijnens:1995xf} approach is an artefact of the contact interaction therein.

We will employ the Dyson-Schwinger equations (DSEs), which are renormalisable functional integral relations amongst
the Green's functions of the QFT. To satisfy the vector and axial-vector Ward-Takahashi identities (WTI) 
we take the rainbow-ladder truncation~\cite{Maris:1997tm,Maris:1999nt}. This is successful in a wide
range of meson~\cite{Maris:1997tm,Maris:1999nt,Maris:1999bh,Jarecke:2002xd,Maris:2002mz}
and baryon~\cite{Eichmann:2009qa,Eichmann:2011vu,Eichmann:2011pv,SanchisAlepuz:2011jn} properties. 
For a summary of the LOHVP contribution calculated in the DSE approach, compared with
recent Lattice QCD results, see Ref.~\cite{Goecke:2013fpa}.

\section{Comparison: DSE vs. ENJL}
\label{sec:DSEvsENJL}
First, we note that the DSEs are renormalisable and feature a continuous connection
between the infrared and ultraviolet limit. This contrasts with ENJL which is a
non-renormalisable effective model with a cut-off on the order of $1$~GeV.
Secondly, due to its contact interaction the ENJL model features dressing functions
with trivial momentum dependence. The DSE approach is quite different in this respect; see
Ref.~\cite{Goecke:2012qm} for more details. 

The inverse quark propagator is 
\begin{align}
  S^{-1}(p) = Z_f^{-1}(p^2)\,(-i\, \sh{p}+M(p^2))\,,
\end{align}
with mass function $M(p^2)$ and wave-function renormalisation $Z_f(p^2)$. In the DSE approach
both feature a momentum dependence that connects the perturbative ultraviolet limit with the
non-perturbative deep infrared. The ENJL approach~\cite{Bijnens:1995xf,Bijnens:1994ey} is a low energy effective theory
with a contact interaction and so we have $Z_f=1$ and a constituent quark mass $M(p^2)\approx 300$~MeV.
In the DSE approach $Z_f(p^2)<1$ leading to a suppression of the propagator that is compensated
by a comparable enhancement in the quark-photon vertex as constrained by the Ward-Takahashi
identity (WTI)
\begin{align}
	  \i P_\mu \Gamma_\mu(P,k) &= S^{-1}(k_-) - S^{-1}(k_+)\;.
\end{align}
That the DSE mass function connects the current quark mass (a few MeV) to the
constituent-like mass (a few hundred MeV) suggests that on average a lighter
than constituent quark mass is probed, typically of order $200$~MeV. This is the main reason
for the larger relevance of this diagram in our approach.

The quark-photon vertex is more complicated. Both approaches satisfy the WTI and in addition feature
dynamical vector meson poles. However, in the ENJL approach the contact interaction 
decouples loop integral corrections and the bubble sum may be determined
from a geometric series. The vertex has the form
\begin{align}
  \Gamma_\mu^\mathrm{ENJL} = \gamma_\mu - \gamma^T_\mu \frac{Q^2}{Q^2+M_V^2}\,,
  \label{eqn:NJLVertexLT}
\end{align}
which contains the bare vertex $\gamma_\mu$ and the leading transverse structure
$\gamma_\mu^\T=(\delta_{\mu\nu}-Q_\mu Q_\nu/Q^2)\gamma_\nu$. The dressing of this transverse part is given here 
in the VMD limit of the ENJL model \cite{Bijnens:1994ey} where
for two flavours $M_V$ is identified with the $\rho$-mass. Using the transversality of the hadronic photon four-point
function with respect to its photon legs, this vertex can be reduced to
$\gamma_\mu M_V^2/(Q^2+M_V^2)$ which highlights the strong suppression induced by the
VMD contribution to the vertex.

In the DSE approach, the interaction type is non-contact and the vertex is decomposed into
twelve tensor components
\begin{align}
  \Gamma_\mu(Q,k) = \sum_{i=1}^{4} \lambda^{(i)}(Q,k)L_\mu^{(i)} + \sum_{i=1}^{8} \tau^{(i)}(Q,k)T_\mu^{(i)} \,,
  \label{eqn:VertexDecomposition}
\end{align}
where $k$, $Q$ are the relative and quark momentum and the total photon momentum, respectively.
The vector meson bound state appears dynamically in the transverse vertex structure. A simple fit to the 
numerical results of the quark-photon vertex has been provided~\cite{Maris:1999bh}
\begin{align}
  \Gamma_\mu(Q,k) \simeq\Gamma_\mu^\mathrm{BC}\!-\! 
          \gamma_\mu^\T \frac{\omega^4N_V}{\omega^4+k^4}
	  \frac{f_V}{M_V}\frac{Q^2}{Q^2+M_V^2}\, e^{-\alpha(Q^2+M_V^2)},
  \label{eqn:QEDVertexFitToLeadingTransverse}
\end{align}
which consists of the Ball-Chiu part, $\Gamma^{\textrm{BC}}_\mu$, and the leading transverse structure
corresponding to $T_\mu^1=\gamma_\mu^\T$.
Good agreement with the numerical solution is found for the  parameters$\omega = 0.66\,\mbox{GeV}$, 
$\alpha=0.15$ and $N_V f_V/M_V = 0.152$.
Note that, as in the ENJL model, we have in Eq.~(\ref{eqn:QEDVertexFitToLeadingTransverse})
a part that is given via the WTI (the BC vertex) and a transverse part.

We immediately see a large difference between the transverse parts of the two quark-photon vertices.
In the ENJL approach there is only a dependence on $Q$, the total momentum of the photon, whereas
in the DSE approach the relative quark momentum $k$ is also a parameter. By comparison, one may \emph{restore}
this dependence on the relative quark momentum by introducing the function
\begin{align}
  f(k^2) = \frac{\omega^4}{k^4+\omega^4}\,.
\end{align}
Similarly, one could simulate the impact of neglecting the relative quark momentum in the DSE
approach by setting $k=0$ explicitly. These two cases are shown in Table~\ref{tab:1BCandTransverseVertexResults}.

\begin{table}[h]
\centering
\begin{tabular}{l|c||l|c}
\multicolumn{2}{c||}{Without relative momentum} & \multicolumn{2}{c}{With relative momentum}  \\
\hline
$\gamma_{\mu\,\textrm{ENJL}}^T$ & 43 & $\gamma_{\mu\,\textrm{ENJL}}^T f(k^2)$ & 103 \\
$\gamma_{\mu\,\textrm{DSE,fit}}^T(k=0)$ & 43 & $\gamma_{\mu\,\textrm{DSE,fit}}^T$ & 105 \\
$\gamma_{\mu\,\textrm{DSE,calc}}^T(k=0)$ & 41 & $\gamma_{\mu\,\textrm{DSE,calc}}^T$ & 96 \\
\end{tabular}
\caption{Impact of restricting/restoring the relative quark momentum in the leading transverse part of the 
quark-photon vertex for the ENJL model, the DSE fit of Eq.~(\ref{eqn:QEDVertexFitToLeadingTransverse})
and the calculated quark-photon vertex (DSE,calc), on the quark-loop contribution to $a_\mu$. Units are $\times10^{-11}$, and we use two light-quark flavours.}\label{tab:1BCandTransverseVertexResults}
\end{table}

The results clearly show that restricting the relative quark momentum to be identically zero leads to
a significant suppression of the contribution, from $\sim 100$~MeV down to $\sim 40$~MeV.

Thus, the combination of both a dynamical quark mass and a quark-photon vertex with realistic momentum
dependence yields enhancement of the quark-loop contribution to hadronic light-by-light scattering in the
muon $g-2$ as compared with other effective models such as ENJL.

\section{Conclusions}

We presented a summary of our comparison between the DSE approach and that of the
ENJL model with regards to the quark-loop contribution to hadronic light-by-light
scattering in the muon $g-2$~\cite{Goecke:2012qm}. There are important differences
due to momentum dependence.

For the quark mass function, the connection between light current quark masses
at large momenta and heavy constituent quark-masses at small momenta entails from
the mean-value theorem that some average mass in-between is probed. This average mass 
is of the order $200$~MeV, far lighter than typical values considered in say the ENJL
model. Note that this does not imply a constituent quark mass of $\sim200$~MeV since it
is a merely an integrand-weighted average.

In the quark-photon vertex, we found that restricting the relative quark-momentum dependence
to be zero, a natural consequence in the ENJL model due to the contact interaction,
yields important quantitative differences as to the impact of dynamical vector meson
poles. The suppression of the quark-loop reported in the ENJL model is an artefact
of this momentum restriction, and is almost completed mitigated within the DSE approach
due to the full momentum dependence of the vertex considered therein.

Thus, we conclude that the standard value $a_\mu^{LBL}= 105 (26) \times [10^{-11}]$ used in current evaluations of
the anomalous magnetic moment of the muon \cite{Hagiwara:2011af,Benayoun:2012wc} may be too small concerning its central value 
and is probably much too optimistic in its error estimate.

\section*{Acknowledgements}
RW wishes to thank the organisers of excited QCD. This work was supported by the DFG under contract FI 970/8-1 and the Austrian Science Fund FWF under project M1333-N16.


\begin{thebibliography}{99}
\bibitem{Goecke:2012qm}
  T.~Goecke, C.~S.~Fischer and R.~Williams,
  Phys.\ Rev.\ D {\bf 87} (2013) 034013
  [arXiv:1210.1759 [hep-ph]].


\bibitem{Goecke:2011pe}
  T.~Goecke, C.~S.~Fischer and R.~Williams,
  Phys.\ Lett.\ B {\bf 704} (2011) 211
  [arXiv:1107.2588 [hep-ph]].

\bibitem{Goecke:2013fpa} 
  T.~Goecke, C.~S.~Fischer and R.~Williams,
  Quark Confinement and the Hadron Spectrum X (Confinement X)  
  arXiv:1302.5252 [hep-ph].

\bibitem{Aoyama:2012wk}
  T.~Aoyama, M.~Hayakawa, T.~Kinoshita and M.~Nio,
  Phys.\ Rev.\ Lett.\  {\bf 109} (2012) 111808
  [arXiv:1205.5370 [hep-ph]].


\bibitem{Czarnecki:2002nt}
  A.~Czarnecki, W.~J.~Marciano and A.~Vainshtein,
  Phys.\ Rev.\ D {\bf 67} (2003) 073006
   [Erratum-ibid.\ D {\bf 73} (2006) 119901]
  [hep-ph/0212229].


\bibitem{Hagiwara:2011af}
  K.~Hagiwara, R.~Liao, A.~D.~Martin, D.~Nomura and T.~Teubner,
  J.\ Phys.\ G {\bf 38} (2011) 085003
  [arXiv:1105.3149 [hep-ph]].


\bibitem{Jegerlehner:2009ry}
  F.~Jegerlehner and A.~Nyffeler,
  Phys.\ Rept.\  {\bf 477} (2009) 1
  [arXiv:0902.3360 [hep-ph]].


\bibitem{Feng:2011zk}
  X.~Feng, K.~Jansen, M.~Petschlies and D.~B.~Renner,
  Phys.\ Rev.\ Lett.\  {\bf 107} (2011) 081802
  [arXiv:1103.4818 [hep-lat]].


\bibitem{Boyle:2011hu}
  P.~Boyle, L.~Del Debbio, E.~Kerrane and J.~Zanotti,
  Phys.\ Rev.\ D {\bf 85} (2012) 074504
  [arXiv:1107.1497 [hep-lat]].


\bibitem{DellaMorte:2011aa}
  M.~Della Morte, B.~Jager, A.~Juttner and H.~Wittig,
  JHEP {\bf 1203} (2012) 055
  [arXiv:1112.2894 [hep-lat]].


\bibitem{Bijnens:1995xf}
  J.~Bijnens, E.~Pallante and J.~Prades,
  Nucl.\ Phys.\ B {\bf 474} (1996) 379
  [hep-ph/9511388].


\bibitem{Hayakawa:1995ps}
  M.~Hayakawa, T.~Kinoshita and A.~I.~Sanda,
  Phys.\ Rev.\ Lett.\  {\bf 75} (1995) 790
  [hep-ph/9503463].


\bibitem{Knecht:2001qf}
  M.~Knecht and A.~Nyffeler,
  Phys.\ Rev.\ D {\bf 65} (2002) 073034
  [hep-ph/0111058].


\bibitem{Melnikov:2003xd}
  K.~Melnikov and A.~Vainshtein,
  Phys.\ Rev.\ D {\bf 70} (2004) 113006
  [hep-ph/0312226].


\bibitem{Dorokhov:2008pw}
  A.~E.~Dorokhov and W.~Broniowski,
  Phys.\ Rev.\ D {\bf 78} (2008) 073011
  [arXiv:0805.0760 [hep-ph]].


\bibitem{Dorokhov:2012qa}
  A.~E.~Dorokhov, A.~E.~Radzhabov and A.~S.~Zhevlakov,
  Eur.\ Phys.\ J.\ C {\bf 72} (2012) 2227
  [arXiv:1204.3729 [hep-ph]].


\bibitem{Greynat:2012ww}
  D.~Greynat and E.~de Rafael,
  JHEP {\bf 1207} (2012) 020
  [arXiv:1204.3029 [hep-ph]].


\bibitem{Cappiello:2010uy}
  L.~Cappiello, O.~Cata and G.~D'Ambrosio,
  Phys.\ Rev.\ D {\bf 83} (2011) 093006
  [arXiv:1009.1161 [hep-ph]].


\bibitem{Fischer:2010iz}
  C.~S.~Fischer, T.~Goecke and R.~Williams,
  Eur.\ Phys.\ J.\ A {\bf 47} (2011) 28
  [arXiv:1009.5297 [hep-ph]].


\bibitem{Goecke:2010if}
  T.~Goecke, C.~S.~Fischer and R.~Williams,
  Phys.\ Rev.\ D {\bf 83} (2011) 094006
   [Erratum-ibid.\ D {\bf 86} (2012) 099901]
  [arXiv:1012.3886 [hep-ph]].


\bibitem{Prades:2009tw}
  J.~Prades, E.~de Rafael and A.~Vainshtein,
  (Advanced series on directions in high energy physics. 20)
  [arXiv:0901.0306 [hep-ph]].


\bibitem{Bennett:2006fi}
  G.~W.~Bennett {\it et al.}  [Muon G-2 Collaboration],
  Phys.\ Rev.\ D {\bf 73} (2006) 072003
  [hep-ex/0602035].


\bibitem{Roberts:2010cj}
  B.~L.~Roberts,
  Chin.\ Phys.\ C {\bf 34} (2010) 741
  [arXiv:1001.2898 [hep-ex]].


\bibitem{Benayoun:2012wc}
  M.~Benayoun, P.~David, L.~DelBuono and F.~Jegerlehner,
  arXiv:1210.7184 [hep-ph].


\bibitem{Maris:1997tm}
  P.~Maris and C.~D.~Roberts,
  Phys.\ Rev.\ C {\bf 56} (1997) 3369
  [nucl-th/9708029].


\bibitem{Maris:1999nt}
  P.~Maris and P.~C.~Tandy,
  Phys.\ Rev.\ C {\bf 60} (1999) 055214
  [nucl-th/9905056].


\bibitem{Maris:1999bh}
  P.~Maris and P.~C.~Tandy,
  Phys.\ Rev.\ C {\bf 61} (2000) 045202
  [nucl-th/9910033].


\bibitem{Jarecke:2002xd}
  D.~Jarecke, P.~Maris and P.~C.~Tandy,
  Phys.\ Rev.\ C {\bf 67} (2003) 035202
  [nucl-th/0208019].


\bibitem{Maris:2002mz}
  P.~Maris and P.~C.~Tandy,
  Phys.\ Rev.\ C {\bf 65} (2002) 045211
  [nucl-th/0201017].


\bibitem{Eichmann:2009qa}
  G.~Eichmann, R.~Alkofer, A.~Krassnigg and D.~Nicmorus,
  Phys.\ Rev.\ Lett.\  {\bf 104} (2010) 201601
  [arXiv:0912.2246 [hep-ph]].


\bibitem{Eichmann:2011vu}
  G.~Eichmann,
  Phys.\ Rev.\ D {\bf 84} (2011) 014014
  [arXiv:1104.4505 [hep-ph]].


\bibitem{Eichmann:2011pv}
  G.~Eichmann and C.~S.~Fischer,
  Eur.\ Phys.\ J.\ A {\bf 48} (2012) 9
  [arXiv:1111.2614 [hep-ph]].


\bibitem{SanchisAlepuz:2011jn}
  H.~Sanchis-Alepuz, G.~Eichmann, S.~Villalba-Chavez and R.~Alkofer,
  Phys.\ Rev.\ D {\bf 84} (2011) 096003
  [arXiv:1109.0199 [hep-ph]].


\bibitem{Bijnens:1994ey}
  J.~Bijnens and J.~Prades,
  Z.\ Phys.\ C {\bf 64} (1994) 475
  [hep-ph/9403233].


\bibitem{Ball:1980ay}
  J.~S.~Ball and T.~-W.~Chiu,
  Phys.\ Rev.\ D {\bf 22} (1980) 2542.

\end{thebibliography}
\end{document}